\documentclass[twocolumn,showpacs,showkeys,preprintnumbers,amsmath,amssymb]{revtex4}
\usepackage{graphicx}
\usepackage{dcolumn}
\usepackage{bm}
\usepackage{subfigure}
\usepackage{caption}
\usepackage{ulem}
\begin{document}

\preprint{To appear in Physical Review E}

\title{Evaluation of Scale-Invariance In Physiological Signals By Means Of Balanced Estimation Of Diffusion Entropy\\}
\author{Wenqing Zhang$^1$}
\author{Lu Qiu$^1$}
\author{Qin Xiao$^1$}
\author{Huijie Yang$^1$}
\email{hjyang@ustc.edu.cn} \altaffiliation {Corresponding author}
\author{Qingjun Zhang$^2$}
\author{Jianyong Wang$^3$}
\address{1 Business School, University of Shanghai for Science and Technology,Shanghai 200093,China\\
2 Analysis and Testing Center, Hebei Polytechnic University, Tangshan 063009, Hebei Province, China\\
3 Department of Physics, Xingtai College, Xingtai 054001, Hebei Province, China}

\date{\today}

\begin{abstract}
By means of the concept of balanced estimation of diffusion entropy we evaluate reliable scale-invariance embedded in different sleep stages and stride records. Segments corresponding to Wake, light sleep, REM, and deep sleep stages are extracted from long-term EEG signals. For each stage the scaling value distributes in a considerable wide range, which tell us that the scaling behavior is subject- and sleep cycle- dependent. The average of the scaling exponent values for wake segments is almost the same with that for REM segments ($\sim 0.8$). Wake and REM stages have significant high value of average scaling exponent, compared with that for light sleep stages ($\sim 0.7$). For the stride series, the original diffusion entropy (DE) and balanced estimation of diffusion entropy (BEDE) give almost the same results for de-trended series. Evolutions of local scaling invariance show that the physiological states change abruptly, though in the experiments great efforts have been done to keep conditions unchanged. Global behaviors of a single physiological signal may lose rich information on physiological states. Methodologically, BEDE can evaluate with considerable precision scale-invariance in very short time series ($\sim 10^2$), while the original DE method sometimes may underestimate scale-invariance exponents or even fail in detecting scale-invariant behavior. The BEDE method is sensitive to trends in time series. Existence of trend may leads to a unreasonable high value of scaling exponent, and consequent mistake conclusions.

\end{abstract}

\maketitle

\section{Introduction}
\label{intro}
Scale-invariance embedded in physiological signals can shed light on mechanisms of dynamical processes occurring in human body, based upon which one can construct theoretical models of the processes and evaluate healthy states of disease suffers \citep{scaleinvariance1,scaleinvariance2}. A typical example is the scaling behaviors in different sleep stages. A cycle of healthy sleep persists typically $1$ to $2$ hours, which constitutes a sequence of sleep stages including wake, light sleep, rapid eye movement (REM), and deep sleep. Little is known about the specific functions of these circadian rhythms, it is believed that the deep and REM sleep states are essential for physical recreation and memory reconsolidation, respectively. Extensive works show that heartbeat dynamics are characterized by long-range correlations and different long-range exponents are found for healthy and patients suffering from disease \citep{diseasehurst}, and for different sleep stages \citep{stagehurst}. In particular, at different sleep stages long-range correlation occurs solely during REM sleep, which is similar while less pronounced to that during wakefulness. Hence, scale-invariance embedded in heartbeat intervals can be used as an monitor of intrinsic neural-autonomic regulation of the circadian rhythms, which may find its potential use in disease diagnosis and therapy. But evaluation of scale-invariance in physiological signals meets two challenges.

Methodologically, variance-based methods, such as Wavelet analysis \citep{wavelet} and De-trended fluctuation approach\citep{dfa}, are widely used in literature to calculate scaling exponents. They can estimate correctly values of scaling exponents for fractional Brownian motions, but incorrectly for Levy walks, and even can not find out a scaling-invariant behavior existing in a Levy flight process due to divergency of the second moment \citep{variancemethod}. A successful complementary method is diffusion entropy analysis. From a stationary time series, one can construct all the possible segments with a specified length. Regarding the length as time duration, each segment can be regarded as a trajectory of a particle starting from original point. The time series is then mapped to an ensemble with the trajectories being realizations of a stochastic motion. From distribution function of displacement one can calculate Shannon entropy, which is called diffusion entropy by Scafetta et al.\citep{variancemethod}. Detailed works prove its powerful in evaluation of scaling exponents for both fractional Brownian and Levy motions\citep{deaapplication}.

Practically, to obtain the probability distribution function, we divide the distribution region of displacement into many bins and reckon the number of displacements occurring in each bin. The occurring probability at a bin is generally approximated with relative frequency, namely, ratio between the occurring number and the total number of realizations, which is perfect for an ensemble with infinite number of realizations. However, a physiological signal is generally very short. Sometimes we can obtain a long time series, but there occur some phase transitions in the measuring duration, which separate the series into short segments with different scaling behaviors, respectively. For example, a typical cycle of healthy sleep contains several thousands of heartbeat intervals, in which a sequence of transitions occur between different sleep stages. Figure 2(a) presents a long-term heartbeat interval series for a healthy subject. The lengths for wake, REM, light sleep, and deep sleep stages distribute in wide regions from $10^1$ to $10^3$ without characteristic lengths, respectively (see materials).

Short length of time series may induce large statistical fluctuations and/or bias to physical quantities such as probability, moment, and entropy \citep{shortbias}. In a recent paper, Bonachela et al. \citep{Bonachela} review the efforts for improved estimators of entropy for small data sets and propose accordingly a balanced estimator that performs well when the data sets are small. In one of our recent papers, we propose a new concept called balanced estimator of diffusion entropy (BEDE) \citep{YangBEDE}, in which the original form of entropy in diffusion entropy analysis is replaced with the balanced estimator of entropy. Calculations show that it gives reliable scaling exponents for short time series with length $\sim 10^2$.

In the present paper, the BEDE method is used to evaluate scaling behaviors in heartbeat series for different sleep stages and stride time series for normal, fast, and slow walkers. Results show that for finite records of physiological signals the current methods in literature may lead to unacceptable errors for scaling exponents and wrong conclusions, while the BEDE approach can provide us a reliable estimation of scaling exponents.

\section{Method and Materials}
\subsection{Diffusion Entropy}
Let us review briefly the concept of diffusion entropy \citep{variancemethod}proposed to detect scale-invariance in stationary series. For a stationary time series, $\xi_1,\xi_2,\cdots,\xi_N$, all the possible segments with specified length $s$ read
\begin{equation}
X_i(s)=\{\xi_i,\xi_{i+1},\cdots,\xi_{i+s-1}\}, i=1,2,\cdots,N-s+1.
\end{equation}
\noindent We regard the length $s$ as time and consequently $X_i(s)$ the $i$th realization of a stochastic process. The total $N-s+1$ realizations form an ensemble of the process. Displacement of the $i$th realization read
\begin{equation}
x_i(s)=\sum\limits_{j=i}^{i+s-1} \xi_j.
\end{equation}

\noindent Dividing the interval the displacements occur into $M(s)$ bins, one can reckon the number of displacements occurring in each bin, denoted with $n(k,s), k=1,2,\cdots,M(s)$. The probability distribution function can be approximated with the relative frequency
\begin{equation}
p(k,s)\sim \hat{p}(k,s)=\frac{n(k,s)}{N-s+1},k=1,2,\cdots,M(s).
\end{equation}

\noindent The consequent naive approximation of Shannon entropy reads

\begin{equation}
S_{DE}(s)\sim S_{DE}^{naive}(s)=-\sum\limits_{k=1}^{M(s)} \hat{p}(k,s) ln [\hat{p}(k,s)].
\end{equation}

Provided the stochastic process behaves scale-invariant, we have
\begin{equation}
\begin{array}{cc}
p(k,s)\sim \frac{1}{s^\delta} F\left(\frac{x_{min}(s)+(k-0.5)\epsilon(s)}{s^\delta}\right),\\
k=1,2,\cdot,M(s),
\end{array}
\end{equation}
\noindent where $\epsilon(s)$ is the size of bin, which is simply selected to be a certain fraction of standard deviation of the initial series. Plugging Eq.(5) into Eq.(4) leads to
\begin{equation}
S_{DE}(s)= -\int_{-\infty}^{+\infty} dy F(y)ln[F(y)]+\delta ln(s)=A+\delta ln(s),
\end{equation}
\noindent As a powerful method, diffusion entropy (DE) has been used to evaluate scaling invariance embedded in time series in diverse fields, such as solar activities \citep{solar}, spectra of complex networks \citep{network}, physiological signals \citep{stride}, DNA sequences\citep{DNA}, geographical phenomena \citep{geograp}, and finance\citep{finance}.

\subsection{Balanced Estimation of Diffusion Entropy}
Unfortunately, extension of Eq.(6) to the naive approximation of diffusion entropy is a nontrivial step, i.e., generally $S_{DE}^{naive}(s) \ne A+\delta ln(s)$ \citep{shortbias}. Defining relative error, $\mu(k,s)\equiv \frac{\hat{p}(k,s)-p(k,s)}{p(k,s)}$, after a straightforward computation we have,
\begin{equation}
S_{DE}(s)= S_{DE}^{naive}+\frac{M(s)-1}{2(N-s+1)}+O[M(s)].
\end{equation}
\noindent
\noindent The leading order of error, $\frac{M(s)-1}{N-s+1}$, vanishes as $N-s\rightarrow \infty$, while it may become unacceptable large when $N-s$ is finite. Especially, for short time series the linear relation in Eq.(6) will be distorted completely and one can not find scaling invariant behavior. In the naive approximation of diffusion entropy there exist simultaneously statistical error (variance) and systematical error (bias). That is, $\hat{p}(k,s)ln[\hat{p}(k,s)]$ is not an acceptable estimation of its corresponding term $p(k,s)ln[p(k,s)]$. Hence, our task is to find a new estimation of $p(k,s)ln[p(k,s)]$, denoted with $\hat{S}_{DE}[n(k,s)]$, which make combination of variance and bias minimum. Here, we ignore correlations between $n(k,s),k=1,2,\cdots,M(s)$.

We employ the solution proposed by Bonachela et al. \citep{Bonachela}. Mathematically, this problem can be formulated as,
\begin{equation}
\begin{array}{l}
\frac{\partial \Delta^2(k,s)}{\partial \hat{S}_{DE}[n(k,s)]}=0,\\
\Delta^2(k,s)=\int_{0}^{1}\left[\Delta_{bias}^2(k,s)+\Delta_{stat}^2(k,s)\right]w[p(k,s)]dp(k,s),
\end{array}
\end{equation}
\noindent where
\begin{equation}
\begin{array}{l}
\Delta_{bias}^2(k,s)=\left(p(k,s)ln[p(k,s)]-\left<\hat{S}_{DE}[n(k,s)]\right>\right)^2,\\
\Delta_{stat}^2(k,s)=\left<\left(\hat{S}_{DE}[n(k,s)]-\left<\hat{S}_{DE}[n(k,s)]\right>\right)^2\right>,
\end{array}
\end{equation}
\noindent are bias and variance, respectively. And $w[p(k,s)]$ is weight function depending on specific problem. Generally, we set $w[p(k,s)]=1$ due to lack of extra knowledge. The average $<.>$ is conducted by using binomial distribution function,
\begin{equation}
\begin{array}{l}
 P_{n(j,s)} [p(j,s)] = \frac{[N - s + 1]!}{n(j,s)![N - s + 1 -
n(j,s)]!}\times \\
 {\begin{array}{*{20}c}
 {{\begin{array}{*{20}c}
 \hfill \\
\end{array} }} \hfill \\
\end{array} }{\begin{array}{*{20}c}
 \hfill \\
\end{array} }\left[ {p(j,s)} \right]^{n(j,s)} \cdot [1 - p(j,s)]^{N - s + 1
- n(j,s)}. \\
 \end{array}
\end{equation}

\noindent A simple computation leads to \citep{YangBEDE},
\begin{equation}
\hat{S}_{DE}[n(k,s)]=\frac{n(j,s)+1}{N-s+3}\cdot \sum\limits_{k=n(j,s)+2}^{N-s+3}{\frac{1}{k}}.
\end{equation}
\noindent Consequently, a proper estimation of $S_{DE}(s)$ reads,
\begin{equation}
\hat{S}_{DE}(s)=\frac{1}{N-s+3}\sum\limits_{j=1}^{M(s)}
[n(j,s)+1]\cdot \sum\limits_{k=n(j,s)+2}^{N-s+3} \frac{1}{k},
\end{equation}
\noindent called Balanced Estimator of Diffusion Entropy (BEDE).

\subsection{Materials}

We consider long-term EEG signals for a total of $16$ male subjects aged from $32$ to $ 56$ (mean age $43$), with weights from $89$ to $152kg$ (mean weight $119kg$) \citep{sleeprecord}. Each EEG record persists averagely $7.5$ hours annotated with sleep staging and apnea information. Each annotation applies to thirty seconds following it. Sleep stages are divided into four stages, namely, deep sleep, light sleep, REM sleep, and wake phase, which are determined by using visual evaluation of electrophysiological recordings of brain activity.

We consider also stride series for a total of $10$ young healthy volunteers,  denoted with $si01,si02,\cdots,si10$ \citep{striderecord}. Healthy here refers to that the participants have not history of any neuromusucular, respiratory, or cardiovascular disorders and are taking no medication. Age distributes in $18$ to $29$ years. Average age is $21.7$ years. Height and weight center at $177cm$ and $71.8kg$, with standard deviations $8cm$ and $10.7kg$, respectively. All the subjects walk continuously on level ground around an obstacle free, long (either $225m$ or $400m$), approximately oval path. The stride interval is measured by using ultra-thin, force sensitive switches taped inside one shoe. Each subject walks with four trials, including slow, normal, fast, and metro-regulated. For the slow, normal, and fast trials the mean stride intervals are $1.3\pm 0.2m, 1.1\pm 0.1m$ and $1.0\pm 0.1m$, and the mean walking rates are $1.0\pm 0.2m/s,1.4\pm 0.1m/s$ and $1.7\pm 0.1m/s$, respectively.

Scale-invariance embedded in heartbeat/stride interval series, denoted with $\{y^O_1,y^O_2,\cdots,y^O_N\}$, are evaluated. The key step in conducting BEDE is to guarantee the considered series being stationary \citep{trend}. The centered moving average method \citep{cma} is employed to obtain trend of a time series, namely, from the original series we calculate its trend $y^T$, where the elements read,
\begin{equation}
\begin{array}{l}
y^T_{i}=\frac{1}{s}\cdot\sum_{j=-[(s+1)/2]+1}^{[s/2]}y^O_{i+j},\\
 i=[(s+1)/2],[(s+1)/2]+1,\cdots,N-[s/2].
\end{array}
\end{equation}
\noindent Herein, the size of moving window is identical with $s$ in $\hat{S}_{DE}(s)$. The consequent de-trended series $y^D$ read,
\begin{equation}
y^D_{i}=y^O_i-y^T_i, i=[(s+1)/2],[(s+1)/2]+1,\cdots,N-[s/2].
\end{equation}

Fractional Brownian motions are generated to investigate the performance of centered moving average. Figure 1(a)-(c) present results for three series with Hurst exponents $H=0.9,0.7$ and $0.3$, as typical examples. DE and BEDE methods are used to estimate $\delta$ values for the original signals and the corresponding de-trended series. The curves show that from the de-trended series BEDE can obtain almost the same values of $\delta$ compared with that from the original series, namely, in the BEDE method the centered moving average does not introduce artificial characteristics. From the original series, BEDE can obtain correctly values of $\delta$, while DE method can not find scaling invariance in signals with larger values of $H$ (the curves bend down, especially for signals with large $H$). Fig.1(d) show the variance and bias (mean) of estimated exponents for de-trended series by using DE and BEDE, respectively. The average is conducted over $1000$ series with length $300$ for each $H$. One can find that the BEDE method can estimate $\delta$ for signals with $0<\delta<1$ without bias and with higher precision, while DE method underestimates $\delta$ up to $10\%$ and has comparatively lower precision.

\begin{figure}
\scalebox{1.3}[1.3]{\includegraphics{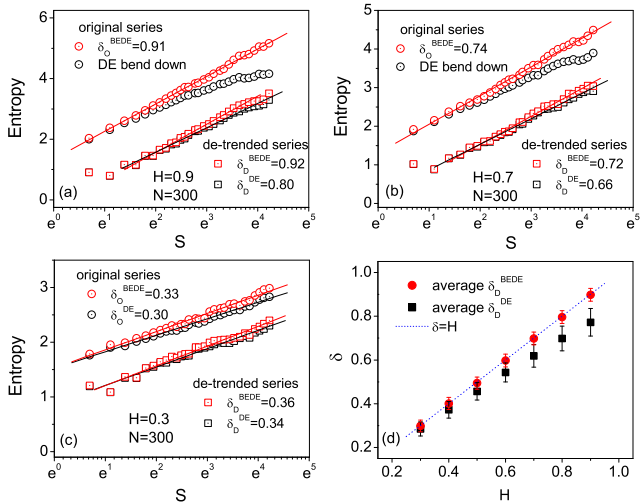}}
\caption{\label{fig:epsart} (Color online) Performance of centered moving average. The series are fractional Brownian motions with Length $300$. (a)-(c) Results for three series with $H=0.9,0.7$, and $0.3$, respectively. BEDE can accurately estimate $\delta$ from the original and de-trended series. DE curve for the original series bend down. DE can detect scale-invariance in the de-trended series. Confident intervals for estimated values are all less than $0.02$. (d) Bias and deviation of the DE and BEDE estimations for de-trended series. BEDE can estimate scaling exponent without bias and with higher precision, while DE underestimates scaling exponent up to $10\%$ with lower precision.}
\end{figure}

\begin{figure}
\scalebox{1.2}[1.2]{\includegraphics{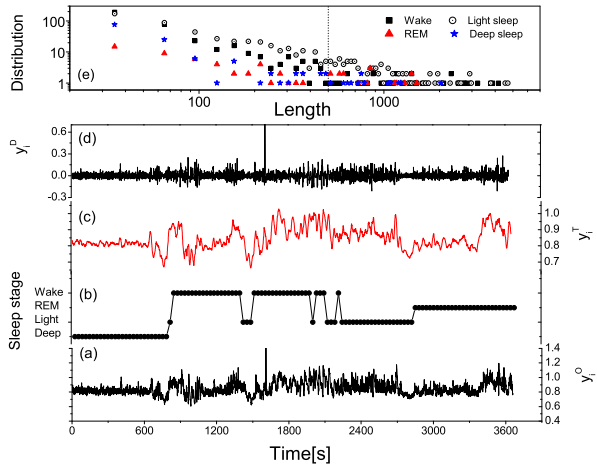}}
\caption{\label{fig:epsart} (Color online) EEG records. (a) Heartbeat interval series for subject numbered $59$, part is shown as an example. (b) Sleep stages annotated by visual evaluation of electrophysiological recordings of brain activity.(c) Trend extracted from the original series in (a). Trend for $s=21$ is shown as an example. (d) De-trended series for $s=21$ as an example. (e) Lengths for wake, light sleep, REM, and deep sleep stages distribute in $10^1\sim 10^3$ (without characteristic lengths).}
\end{figure}

\begin{figure}
\scalebox{1.4}[1.5]{\includegraphics{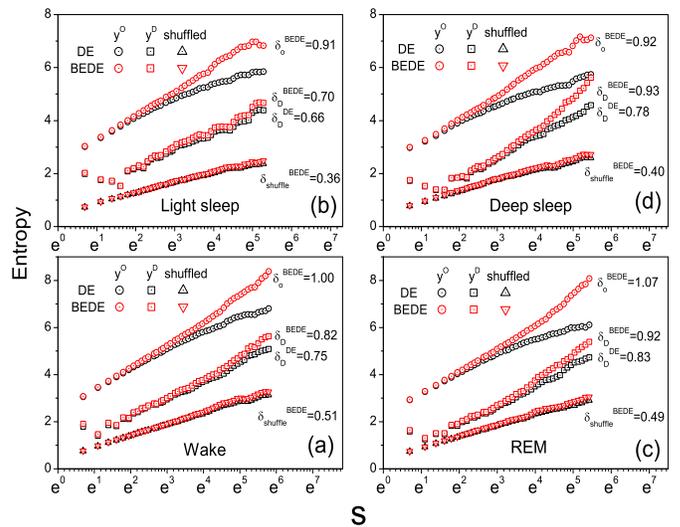}}
\caption{\label{fig:epsart} (Color online) Scaling behaviors in a sleep cycle of the subject numbered $59$. (a)-(d) Scaling behaviors of the segments $7290s-8520s$(wake), $8520s-9270s$(light sleep), $9270s-10110s$(REM), and $ 6450s-7290s$(deep sleep), respectively. Confident intervals for estimated values are all less than $0.02$.}
\end{figure}

\begin{figure}
\scalebox{1.4}[1.4]{\includegraphics{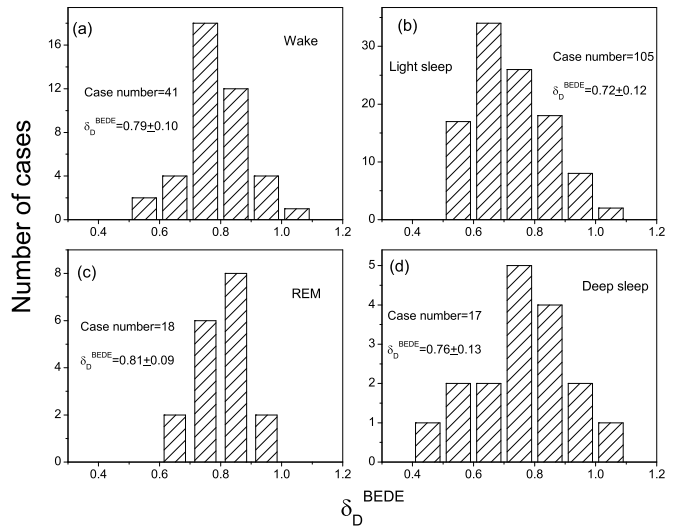}}
\caption{\label{fig:epsart} (Color online) Distribution of scaling exponents for different sleep stages. From all the subjects the segments corresponding to wake, light sleep, REM, and deep sleep are extracted, respectively. Segments whose lengths are less than $500$ are discarded. (a)-(d) Scaling exponent distributions for wake, light sleep, REM, and deep sleep, respectively. The average values for REM and wake are significantly large ($\sim 0.8$) compared with that for light sleep ($\sim 0.7$).}
\end{figure}

\section{Results}
\subsection{Scaling Behaviors for Sleep Stages}
Figure 2(a) shows the heartbeat interval series for subject numbered $59$ as a typical record (part). One can find that transition between different sleep stages occurs frequently, as annotated in Fig.2(b). There exists a complicated and significant trend (see Fig.2(c), in which trend for $s=21$ is shown as an example). The corresponding de-trended series is depicted in Fig.2(d). From all the records one can find that most of lengths for wake, light sleep, REM, and deep sleep stages distribute in $10^1\sim 10^3$ (without characteristic lengths), as presented in Fig.2(e).

As a typical example, we show in Fig.3(a)-(d) results for the segments $7290s-8520s$(wake), $8520s-9270s$(light sleep), $9270s-10110s$(REM), and $ 6450s-7290s$(deep sleep), which forms a sleep cycle of the subject numbered $59$, respectively. One can find that the curves for DE results bend down with the increase of scale, while this trend are corrected by BEDE to straight lines in a considerable range of scale.  What is more, though BEDE can detect successfully scaling behaviors in the original series, the estimated values of scaling exponents are significantly large compared with that for the corresponding de-trended series. From the original series the estimations for wake, light sleep, REM, and deep sleep are $1.00,0.91,1.07,0.92$, while that from the corresponding de-trended series are $0.82,0.70,0.92,0.93$. For the de-trended series, DE gives underestimates of the exponents up to $10\%$. Hence, to obtain reliable scaling behaviors for the different sleep stages we must conduct de-trend procedure and use BEDE instead of DE. Results for shuffled de-trended series tell us that the scaling behaviors come from non-trivial patterns in series rather than distribution of elements in series.

From all the subjects we extract the segments corresponding to wake, light sleep, REM, and deep sleep, whose lengths are larger than $500$ (about a duration of $450s$), the number of which are $41,105,18$ and $17$, respectively. The distribution behaviors of the values for scaling exponent are presented in Fig.4(a)-(d). One can find that the average value and standard deviation of scaling exponent for REM are almost the same with that for wake, but the distribution for the former one is much sharper than that for the later one. The average value and standard deviation for light sleep are almost identical with that for deep sleep, while the detailed shapes for the distributions are different completely. Obviously, to confirm if the differences in distributions originate from intrinsic behaviors or just from statistical errors, requires collection of a large amount of cases. What is more, the average values for REM and wake are significantly large ($\sim 0.8$) compared with that for light sleep ($\sim 0.7$).

\begin{figure}
\scalebox{1.3}[1.3]{\includegraphics{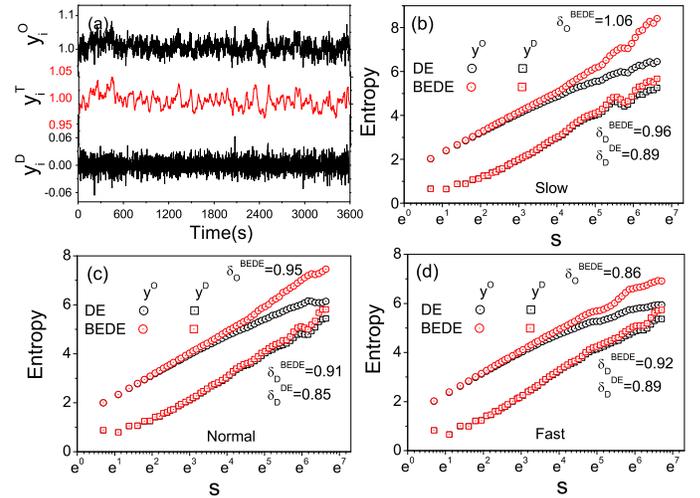}}
\caption{\label{fig:epsart} (Color online) Scale-invariance of stride series. (a) Stride interval series from normal walk record of the volunteer numbered $si01$. And the corresponding trend and de-trended series. Trend and de-trend series for s=21 are shown as an example. (b)-(d) Scaling behaviors in slow, normal, and fast walk series. For the de-trended series, the estimated values by using DE are very close to (generally smaller than) that by using BEDE. Confident intervals for estimated values are all less than $0.02$. }
\end{figure}

\begin{figure}
\scalebox{1.4}[1.4]{\includegraphics{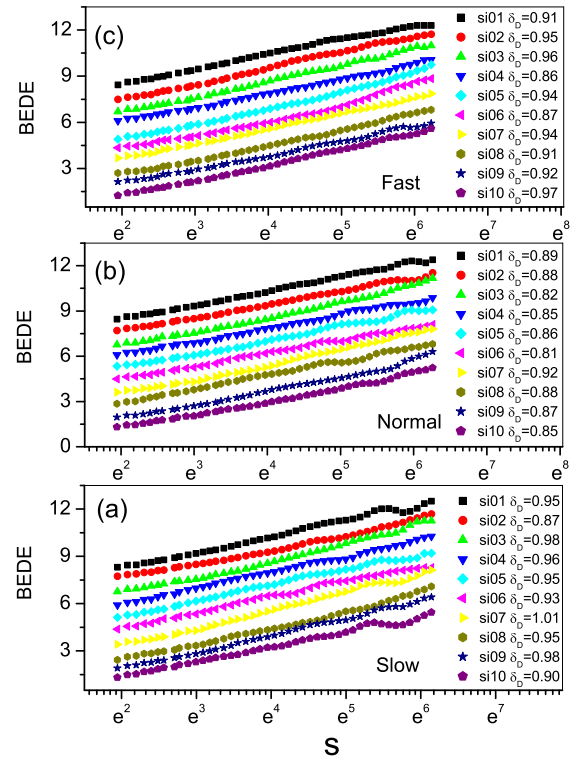}}
\caption{\label{fig:epsart} (Color online) The BEDE results of de-trended series for all the ten volunteers. (a)-(c) Scaling behaviors embedded in slow, normal, and fast series (de-trended). Confident intervals for estimated values are all less than $0.02$.}
\end{figure}

\begin{figure}
\scalebox{1.4}[1.4]{\includegraphics{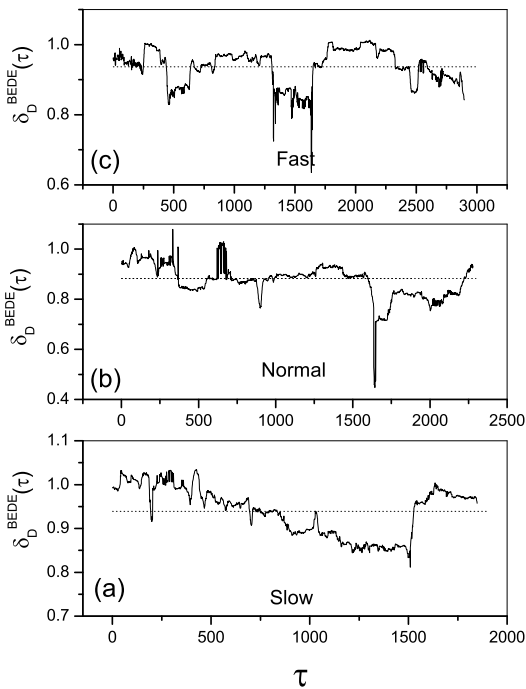}}
\caption{\label{fig:epsart} (Color online) Evolution of scaling behaviors in stride records. (a)-(c) Scaling behaviors embedded in slow, normal, and fast walk records of the volunteer numbered $si01$. The scaling exponents oscillate abruptly in wide ranges. $\Delta\tau=700 $}
\end{figure}

\begin{figure}
\scalebox{1.6}[1.6]{\includegraphics{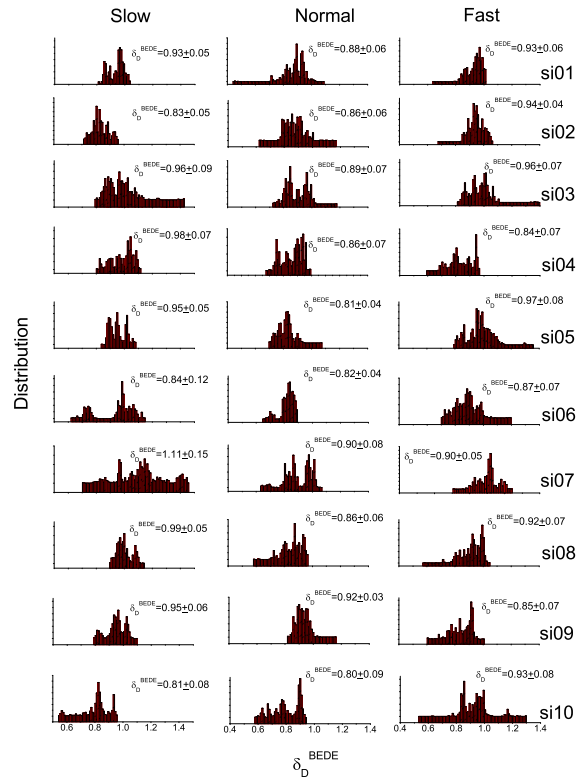}}
\caption{\label{fig:epsart} (Color online) Distribution details of the evolutionary scaling exponents for slow, normal , and fast walking series of all the ten volunteers. Details of the distributions are different completely. $\Delta\tau=700 $.}
\end{figure}

\subsection{Scaling Behaviors for Stride Series}
From stride records, one can calculate the corresponding stride interval series, trends, and de-trended series. As an example, see in Fig.5(a) results from normal walk record of the volunteer numbered $si01$. For slow, normal, and fast walks the lengths of the series are $3304,3371,3395$, respectively, which are significantly larger than that of the sleep stages. For the original series with the increase of scale the DE curves bend down, i.e., scaling behavior can not be detected successfully, while the BEDE curves are almost perfect straight lines in considerable wide scales, as shown in Fig.5(b)-(d). However, the estimated values of scaling exponent from the original series maybe unacceptable large compared with that from the corresponding de-trended series (e.g., the difference for slow record is $0.10$). Consequently, de-trend procedure is the key step to obtain reliable estimations of scaling exponent. What is more, the estimated values by using DE are very close to (generally smaller than) that by using BEDE, and this conclusion stands for all the ten volunteers (not presented).

The BEDE results of de-trended series, $y^{D}$, for all the ten volunteers are shown in Fig.6. The BEDE curves are almost perfect straight lines in considerable width of scale, and the estimations of scaling exponents, namely, slopes of the curves, are also presented.

An interesting question is if the physiological states of the volunteers keep unchanged in the walking experiments. Sliding a window along a de-trended series, at the $\tau$'th step the window covers segment $y_{\tau}^{D},y_{\tau+1}^{D},\cdots,y_{\tau+\Delta\tau-1}^D$, where $\Delta\tau$ is size of the window. Representing the local scaling behavior at step $\tau$ by the scaling exponent for the segment covered by the window, denoted with $\delta_D^{BEDE}(\tau)$, the successive values of $\delta_D^{BEDE}(\tau),\tau=1,2,\cdots,N-\Delta\tau+1$ present the evolution of scaling behavior of the considered series. As a typical result, Fig.7(a)-(c) provide evolutions of scaling behaviors embedded in slow, normal, and fast walk records of the volunteer numbered $si01$. Fig.8 provides the distribution details, mean and standard values of the evolutionary scaling exponents for slow, normal , and fast walking series of all the volunteers. Unexpectedly, the scaling exponents oscillate abruptly in wide ranges and the details of local scaling exponent distributions are different completely. $\Delta\tau $ is chosen to be $700$ in the calculations.

\section{Conclusions}
Scale-invariance in physiological signal records attracts special attentions for its importance in understanding and consequent modeling mechanisms of physiological phenomena and its potential usage in diagnoses and therapy. But evaluation of scale-invariance in physiological signal is a non-trivial task. Theoretically, variance-based methods may lead to a failure evaluation of scaling behavior. Practically, a physiological signal generally is itself very short ($\sim 10^2$), or separated by frequent occurrences of phase transitions into short segments. In literature, persistent efforts have been done to find reliable methods to evaluate scale-invariance in short time series \cite{shorttime}.

In the present paper, we extract wake, light sleep, REM, and deep sleep segments (length $>500$ for reliable estimation of scaling exponent) from long-term EEG signals. By means of the concept of balanced estimation of diffusion entropy (BEDE), we estimate the scaling exponents of de-trended series constructed from the segments. It is found that for each stage the scaling value distributes in a considerable wide range, i.e., the scaling behavior is subject- and sleep cycle- dependent. Statistically, the average of the scaling exponent values for wake segments is almost the same with that for REM segments ($\sim 0.8$), while the average of the scaling exponent values for light sleep segments is about $\sim 0.7$.

For the stride series, because the series are long enough ($3000-4000$), the original diffusion entropy (DE) and balanced estimation of diffusion entropy (BEDE) give almost the same results for de-trended series. But from the evolutions of local scaling invariance one can find that the physiological states change abruptly, though in the experiments great efforts have been done to keep conditions unchanged. Hence, global behaviors of a single physiological signal may lose rich information on physiological states.

Comparison of the results of DE and BEDE, one can find that BEDE can evaluate with considerable precision scale-invariance in very short time series ($\sim 10^2$). The original DE method sometimes underestimates values of scaling exponent, or even can not detect the scaling-behavior due to the bias of bending down. The two methods (BEDE and DE) are all sensitive to trends in time series. In BEDE concept, existence of trend may leads to a unreasonable high value of scaling exponent, which may lead to mistake conclusions. Hence, balanced estimation of diffusion entropy (BEDE) is a preferential candidate to evaluate correctly and precisely scale-invariance in short time series, provided that the time series is de-trended properly.

\begin{center}
\textbf{Acknowledgements}
\end{center}
The work is supported by the National Science Foundation of China
under Grant Nos. 10975099 and 50974052, the Program for Professor of
Special Appointment (Eastern Scholar) at Shanghai Institutions of
Higher Learning, the Innovation Program of Shanghai Municipal Education Commission under Grant No.13YZ072, and the Shanghai leading discipline project under grant No.S30501. One of the authors (W. Zhang) thank the support of the Innovation Fund Project For Graduate Student of Shanghai under Grant No. JWCXSL1102. We thank the reviewers for their stimulating and
constructive comments and suggestions.

\end{document}